\documentclass[journal]{IEEEtran}

\usepackage{xcolor}
\usepackage{color}
\usepackage{soul}

\ifCLASSINFOpdf
    \usepackage[pdftex]{graphicx}
    \usepackage{booktabs} 
    \usepackage{longtable}
    \usepackage{mathtools}
    \graphicspath{{./figs/}}
    \DeclareGraphicsExtensions{.pdf,.jpeg,.png}

\else
\fi
\usepackage[mathlines,switch]{lineno}

\begin{document}

%
\title{Modeling Buried Object Brightness and Visibility for Ground Penetrating Radar}
%
%
%

\author{Garrett~A.~Stevenson, 
        Jason~Wilson, 
        Brian~M.~Worthmann, 
        and~Wlamir~Xavier
\thanks{G.A. Stevenson (stevenson32@llnl.gov), and Dr. B. Worthmann are with the Computational Engineering Division at Lawrence Livermore National Laboratory.}
\thanks{Dr. J. Wilson is chair of the Department
of Mathematics and Computer Science at Biola University and Director of the Biola Quantitative Consulting Center. Dr. W. Xavier is an Associate Professor and member of the Biola Quantitative Consulting Center, Biola University, La Mirada,
CA, 90639.}
}

%
%

\markboth{IEEE TRANSACTIONS ON GEOSCIENCE AND REMOTE SENSING}%
{Shell \MakeLowercase{\textit{et al.}}: Bare Demo of IEEEtran.cls for IEEE Journals}
%



\maketitle

\begin{abstract}
	Comparing the observed brightness of various buried objects is a straightforward way to characterize the performance of a ground penetrating radar (GPR) system. However, a limitation arises. A simple comparison of buried object brightness values does not disentangle the effects of the GPR system itself from the system's operating environment and the objects being observed. Therefore, with brightness values exhibiting an unknown synthesis of systemic, environmental, and object factors, GPR system analysis becomes a convoluted affair. In this work, we use an experimentally collected dataset of over 25,000 object observations from five different multi-static radar arrays to develop models of buried object brightness and control for these various effects. Our modeling efforts provide a means for quantifying the relative brightness of GPR systems, the objects they detect, and the physical properties of those objects which influence observed brightness. To evaluate the models' performance on new object observations, we repeatedly simulate fitting them to half the dataset and predicting the observed brightness values of the unseen half. Additionally, we introduce a method for estimating the probability that individual observations constitute a visible object, which aids in failure analysis, performance characterization, and dataset cleaning.
\end{abstract}

\begin{IEEEkeywords}
ground penetrating radar, tomographic imaging, statistical modeling, explosive hazard detection, buried object brightness, radar array comparison
\end{IEEEkeywords}

%
\IEEEpeerreviewmaketitle

\section{Introduction}

\IEEEPARstart{G}{round} penetrating radar (GPR) has a long-standing history of application to buried object and subsurface anomaly detection \cite{andrews_ralston_tuley_1999,4276906, inproceedings, VOLKOMIRSKAYA2018438,Borrion2019ExperimentalAO}. GPR systems are particularly relevant for the detection of landmines, and the more general field of Explosive Hazard Detection (EHD), leveraging their ability to non-destructively evaluate both metallic and non-metallic buried objects \cite{gros1998survey}. Advances in GPR research include hardware and antenna improvements \cite{TRAVASSOS2018,6757260, doi:10.1002/mop.31724}, as well as software improvements to signal processing and automatic detection algorithms \cite{4276906,BianchiniCiampoli2019,8704991,8081259,7996100,ishitsuka}. Recently, tomographic imaging has been employed to improve detection performance \cite{10.1117/12.186716,osti_1481055,10.1117/12.2519143}.

In this study, we utilize tens of thousands of data points on various objects, at various depths, in multiple sites, observed by multiple antenna arrays over several years. There are three goals in this effort. One, to infer what effect system hardware, environmental factors, and object characteristics have on the observed brightness. Two, to predict the expected brightness of new objects previously unobserved. And three, to use these brightness values to predict the probability that a data point with a given brightness is visible; that is, the image of the object stands out above the background.

Altogether, our models allow for statistically-backed answers to questions such as: ``Which objects are dimmest (\textit{i.e.,} most difficult to detect)?''; ``Which of several radar arrays performs best?''; ``Given an object's characteristics, how bright would it appear at a given depth, for a given array?''; and ``How deep can a particular object be emplaced before it has a 10\% chance of appearing visible?''. This work provides the capability to answer these questions individually within a coherent statistical framework, which is easily extensible to other GPR systems, environments, objects, and use cases.

The remainder of this article is structured as follows: Section \ref{section:objectbrightness} defines the buried object brightness metric, Section \ref{section:Dataset} describes the dataset used in this study, Section \ref{section:ObjectBrightnessModels} introduces brightness models, Section \ref{section:ObjectVisibility} introduces the visibility probability model, Section \ref{section:discussion} discusses the usage of the models and the interpretation of their parameters, and Section \ref{section:Conclusions} provides conclusions, their impact, and directions of future work.

\begin{figure}[h!]
	\centering
	\includegraphics[width=3in]{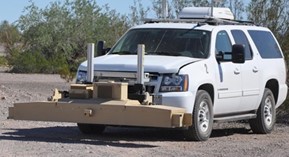}
	\caption{A multi-static radar array used in this study (denoted Array 3) mounted on a ground-based GPR system. \cite{osti_1068301}}
	\label{figure:vehicle}
\end{figure}
\section{Buried Object Brightness}
\label{section:objectbrightness}
To develop the brightness metric, we use tomographic images from processed multi-static radar signals. Figure \ref{figure:vehicle} shows one of the five arrays used in this study. Time series are collected for all 256 transmitter-receiver pairs (16 transmitters, 16 receivers), each with 512 samples in fast time. This collection of 256 time series represents a frame of radar signal data, and a new frame is measured every 2cm in the direction of vehicle travel. Traditionally, GPR antenna arrays operate mono-statically \cite{daniels2006review} and those signals are typically analyzed directly, not through a tomographic imaging algorithm. Therefore, previous analyses of experimentally collected mono-static signals \cite{4276906,8704991,10.1117/12.2519143,4779004} do not directly apply. See \cite{osti_1481055,osti_1068301}, and \cite{SLIpaper} for further GPR system details. In this work, each observation of an object is represented as a single scalar metric, the object's relative brightness. Using a single value for each object observation is simple and sufficient, which allows for easy interpretation of analysis results. 

The time-series are pre-processed to remove the coupling pulse and surface reflection (see \cite{osti_1068301}, Section 4). These signals are then re-aligned to represent a flattened surface and used as input to a computed tomography algorithm, specifically,  plane-to-plane backpropagation \cite{osti_1068301, 10.1117/12.186717}, which creates a three-dimensional GPR image.

These raw images do not compensate for the soil's attenuation effects, nor do they compensate for edge effects near the bounds of the array where fewer transmitter-receiver pairs are available for imaging. Thus, to compensate for these effects, an image standardization method \cite{SLIpaper} is applied which compares each voxel's intensity with an estimate for the clutter intensity at that location. To estimate the clutter intensity, a logarithm of the raw image magnitudes is taken, and then means and standard deviations for each depth and cross-track position are calculated. These means and standard deviations are used to standardize the raw images, using a Z-score methodology \cite{mendenhall2007statistics}. Thus, the standardized image removes the effects of soil attenuation and array edge effects to provide a quantitative metric for each voxel's contrast with the background (\textit{i.e.,} clutter). These values, while fundamentally dimensionless, are described using the term `SLI units', where SLI is an acronym denoting `Standardized Logarithmic Intensity'.

In addition to SLI brightness values, each voxel is assigned a GPS coordinate. A ground truth table containing known buried objects is then used to determine which voxels correspond to the location of a given buried object. The ground truth table contains four Northing and Easting coordinates for each object, representing a tight, pose-variant bounding box. To compensate for GPS error \cite{OXLEY201799}, in addition to the point-spread function of the array, the quadrilateral representing an object is dilated by 25cm, equivalently referred to as an `uncertainty radius'. The use of uncertainty radii is not uncommon in the field of EHD \cite{8704991}, and values in the literature range from 25cm \cite{8704991} to 1m in \cite{osti_1481055}. An inappropriate uncertainty radius has the competing effects of counting clutter as objects (too lenient) or missing objects (too specific). For our GPR and GPS system, 25cm was found by manual inspection of multiple scenarios to be an appropriate uncertainty radius. 

For each buried object, the maximum observed SLI value of all voxels contained within this dilated quadrilateral, which is a maximum projection in depth, is taken to be the scalar metric defining brightness of an object. This value for brightness is the measure of how distinct, if at all, an object appears in GPR data, that is, the peak brightness of an object in the image. In general, the maximum of a random distribution is not a robust metric. However, due to the radar array's point spread function, voxels near the maximum are correlated, owing to the finite pulse width and finite array aperture of the GPR array. Thus, the maximum SLI value near a buried object is indeed a robust measure of that target's brightness. Note that by evaluating only the peak value, considerations involving the object's extent in collected image data (\textit{e.g.,} size, shape, volume \cite{10.1175/JAS-D-16-0102.1}) are not taken into account. This exclusion is intentional, as our modeling efforts focus on isolating the raw performance of GPR systems and not fusing GPR system performance with algorithmic performance.

\begin{figure}
	\centering
	\includegraphics[width=3.4in]{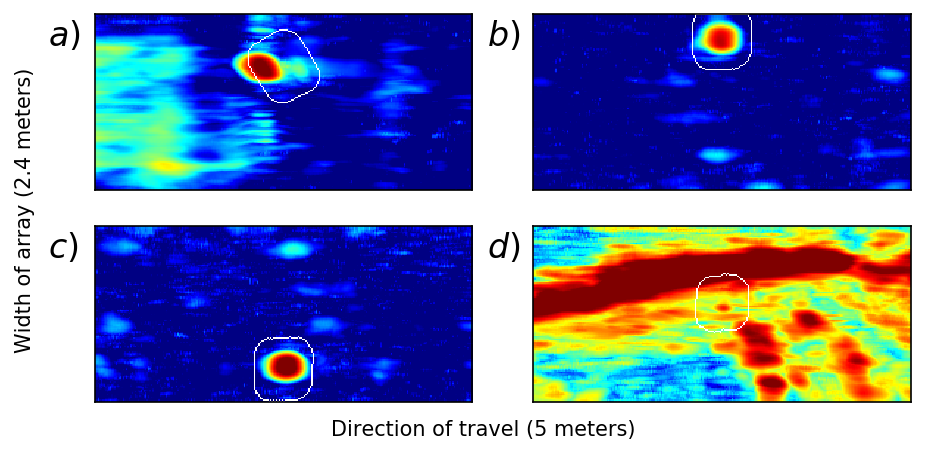}
	\caption{Example object observations in the GPR tomographic image. Vehicle direction is from left to right and a down-track aperture of 5 meters is shown. The white border indicates an uncertainty radius. All 4 images are produced with the same array. Panels \textit{a} and \textit{b} show a low-metallic mine (MINE\_1: shallowest), panel \textit{c} shows a metallic mine (MINE\_2: shallower), and \textit{d} shows a low-metallic mine (MINE\_3: shallow). The maximum brightness within the white border for all 4 panels is, in order, 4.4, 3.8, 4.2, and 4.7 SLI units.}
	\label{figure:1}
\end{figure}

Four object observation examples are displayed in Figure \ref{figure:1}. All four images are max-projected in depth, have the same SLI color scale (1 to 4 SLI units), and come from the same array at different dates/times. Each image is 256 pixels or 5 meters in length and 128 pixels or 2.4 meters across (the swath-width of the array). Panels \textit{a}) and \textit{b}) are views of the same object on different passes (one eastward, the other westward), and the differences in brightness demonstrate some of the inherent variability, 0.56 brightness units in this case, of GPR images. Panel \textit{a}) shows the importance of the uncertainty radius, where GPS error would have caused the brightest part of the object to fall outside the bounding box and therefore lead to an underestimate of that object's maximum brightness value. Panel \textit{d}) shows the opposite problem of an uncertainty-radius and max projection methodology; a very strong surface reflection almost obscures the deeper, fainter object, shown at the center of the outlined region. If an uncertainty radius was not used, or voxels were selected in three dimensions (\textit{e.g.,} controlled for depth), then the brightness value for this object would be lower and therefore more accurate. The observed value of 4.7 here corresponds to the upper edge of the region, where some overlap occurs with the strong surface reflection. Notable examples of the shortcomings in panel \textit{d}) are rare and overwhelmingly, brightness values obtained do appropriately represent the brightness of an object, relative to clutter.

Using a standardized image has several advantages. First, it allows for unbiased comparison of different GPR systems. For example, if two identical systems are considered, but one whose data acquisition system produced values that were an order of magnitude higher than the other, this factor would cancel out at the standardization step. Thus, by comparing different systems, differences in brightness values indicate a stronger separation between foreground (object) and background (clutter). Physically, such a phenomenon could easily occur from differences such as bandwidth, pulse shape or antenna beam width.

Additionally, by allowing the clutter estimate to vary in depth and across the width of an array, the spatially varying incident intensity cancels out in a clutter-limited environment. That is to say, a brighter transmitter or more sensitive receiver does not improve imaging. However, in a noise-limited environment, such as deeper depth, there may remain some depth dependence, as the spatially-varying incident intensity does not cancel out with the spatially homogeneous noise. Further discussion can be found in Section \ref{subsection:depthclutternoise}.

While this work was performed in the tomographic image domain \cite{10.1117/12.186717}, the concept and modeling of buried object brightness is transferable across data modes and application spaces. For example, in the mono-static signal domain \cite{8704991}, the brightness value of an object could be represented by the maximum energy received in a standardized form of the synthetic aperture radar (SAR) signal scan. For brevity, only statistics derived from the tomographic image domain are considered.

Overall, object brightness is quantified through this maximum SLI methodology. While not without shortcomings, it is a simple and intuitive statistic that qualitatively performs well at differentiating bright objects from dim ones. Thus, this statistic is the independent variable chosen for object brightness.

\section{Experimental Dataset}
\label{section:Dataset}
The dataset utilized has a total of 25,643 observations across 449 unique objects. Approximately 23,000 of those observations stem from 415 unique objects emplaced in one Southwestern U.S. testing facility. The remainder come from a second Western U.S. facility. Each observation represents a true experimental result of actual buried objects in a test environment with no simulated conditions other than objects being inert.

\begin{figure}
	\centering
	\includegraphics[width=3.4in]{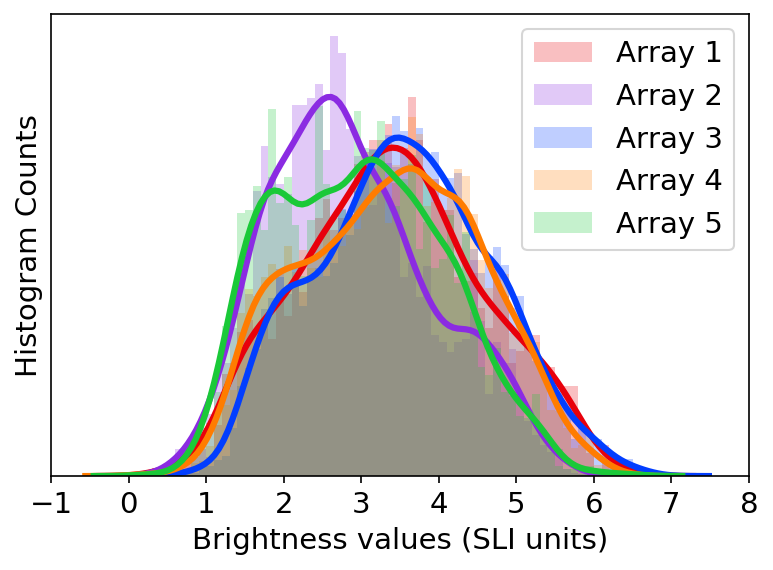}
	\caption{Normalized histogram of object brightness by array.}
	\label{figure:2}
\end{figure}

Figure \ref{figure:2} shows histograms of observed brightness values, controlled by array. The histograms are normalized to unit area (as with probability density functions), though this choice conceals the relative number of objects each array encountered: two arrays collected nearly twice (Array 4) and three times (Array 3) the observations of other arrays, which had approximately 3,000 observations each.

\subsection{GPR System Configuration Variables}
\label{rundata}

Each observation comes from one of 744 collection runs, which are spread across 13 different lanes. Each run has an associated array height and nominal vehicle speed. Array height is treated as a categorical variable, either 30 cm or 45 cm. Vehicle speed is also quantized into a categorical variable as: 5, 10, or 15 kph. Theoretically, based on details of the pre-processing method, these vehicle speeds should not influence object brightness in the observations. The effect of array height was expected to be minimal, with caution of the inverse square law radar signals are subject to. \cite{ander1989test}

\subsection{Object Variables}
\label{objectdata}
Each object is given a unique identifier. An object's identifier has a corresponding name, type, and depth, as well as an associated bounding box area, metallic content, shape, and emplacement date. 

The dataset includes unique object names, each categorized into one of nine general object types. Example object names include MINE\_1, OBJECT\_1, MINE\_2, OBJECT\_2, etc... which obscure specific mine models and object names. While object types include mines\_metallic, mines\_low\_metallic, etc... . These object types were created by hand, to nominally represent a sub-categorization of objects observed in as few categories as is reasonable.

Metallic content is expressed categorically, one of ``metallic'', ``low metallic'', and ``non-metallic''. Object shape is also treated categorically, and is one of: circular, square, rectangular, projectile, irregular, and miscellaneous. Bounding box areas are calculated as the area of the quadrilateral associated with the four bounding box coordinates for each object - note that, due to this definition, a circular object's area is over-estimated by $4/\pi$, or approximately 27\%.

Finally, to test for the possible effect of object degradation over time, the number of days between the object's emplacement and its observation for a given run is also included as a candidate independent variable.

It is important to note the ground truth table through which these object attributes are stored is imperfect. While substantial automated and manual efforts were made to correct errors, undoubtedly some mistakes remain. These mishaps contribute to increased residuals in the models, though the expectation is that in the context of 25,643 observations of unique objects, the effects these mistakes have on the optimal model parameters is mitigated.

\section{Object Brightness modeling}
\label{section:ObjectBrightnessModels}
Modeling object brightness provides a key utility in GPR system analysis. GPR systems are often analyzed in the context of a non-linear detector's performance \cite{4276906,8704991,5650741}, which obscures whether missed detections are due to shortcomings of a system's hardware (an object does not appear at all) or failure of the detection algorithm (an object appears in the data but is not detected). Brightness models fill this void by quantitatively expressing how bright objects appear in the data without bias towards a detection method, which allows for the direct comparison of multiple systems.

A variety of models were considered, including linear regression, random forests, and deep learning. It was determined that linear regression gave a reasonable fit to the data, and was advantageous for the physical interpretation of its parameters, rather than the more ``black box'' approaches. Intelligibility was considered of high value, especially within the context of GPR applied to EHD, as brightness predictions and object visibility have severe connotations.

Specifically, ordinary least squares was utilized with a mixture of categorical and numeric variables. The model parameters associated with categorical variables were held to a constraint of being zero-sum and represented to the regression via dummy encoding. Additionally, a constant term (\textit{i.e.,} y-intercept) was also included in each model and held linearly independent from any numeric variables.

\begin{figure}
	\centering
	\includegraphics[width=3.4in]{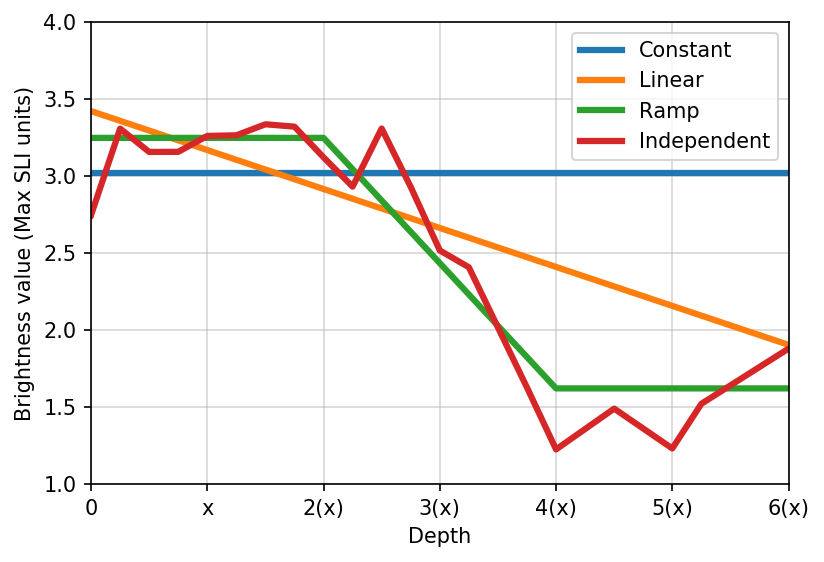}
	\caption{Maximum brightness versus depth for four different depth models: constant, linear, ramp, and independent.}
	\label{figure:depthvariationplot}
\end{figure}
\subsection{Depth Variation}
\label{depthvariation}

Depth is an important parameter for determining object brightness, but due to the standardization procedure which compensates for depth variation in the clutter distribution, it is not immediately obvious how brightness values vary in depth. Thus, different models were developed to model brightness as a function of depth. Each of the four depth models described here includes the variables' object name and array, in addition to depth. Figure \ref{figure:depthvariationplot} illustrates each of these four models' trends in depth assuming an average array and an average object name (\textit{i.e.,} a coefficient of zero for array and object name) and Table \ref{table:depthmodels} displays depth model fitness, where $R^2_{\text{adjusted}}$ (Equation \ref{eq:1}) is used as the figure of merit.
\begin{linenomath*}
\begin{equation}
R^2_{\text{adj.}} = 1 - (1-R^2) \frac{n-1}{n-p}
\label{eq:1}
\end{equation}
\end{linenomath*}
$R^2_{\text{adj.}}$ uses the $R^2$ fitness of a model, the number of data points used $n$, and the number of independent regressors $p$ in the model, to quantify a model's fitness in the context of how many parameters are used.

The \textit{constant} model assumes brightness values are independent of depth, and thus, the constant (intercept) for the model can be considered part of the depth `variation'. This model has just a single free parameter associated with depth, though combined with the array and object name parameters, there are a total of 70 model parameters.

The depth model with the most free parameters, termed \textit{independent}, has a different coefficient for each depth. This is possible because the ground truth table does not have a continuum of depth values, but are rather quantized into 21 different depths. Thus, this model treats each depth as independent from every other depth, practically treating depth as a categorical variable. Note that this model is poorly conditioned, as there is a depth that contains only one object name, and that object name only exists at that depth, leading to a co-linearity, and thus, a condition number tending toward infinity.

A linear model in depth was also considered, which has two free parameters: the slope and the intercept. Finally, a ramp model was devised, which also has two free parameters: the brightness values for upper and lower regions. The location of the ramp vertices, a shallow and medium inflection point were manually chosen based upon the plots in Figure \ref{figure:depthvariationplot}, as well as goodness-of-fit.


\begin{table}[!h]
	\begin{center}
		\begin{tabular}{ccccc}
			\toprule
			Depth	&	RMS of		& $R^2_{\text{adj.}}$		& Condition	& No. of Depth \\
			Model	&	Residuals	&  & Number &  Variables \\
			\midrule
			Constant    &  0.690 &     0.637 &            125.1 &                       1 \\
			Linear      &  0.645 &     0.682 &            125.6 &                       2 \\
			Ramp        &  0.623 &     0.704 &            121.3 &                       2 \\
			Independent &  0.614 &     0.712 &              $\infty$ &                      21 \\
			\bottomrule
		\end{tabular}
	\end{center}
	\caption{Overview of goodness-of-fit for the depth models in Fig. \ref{figure:depthvariationplot}}
	\label{table:depthmodels}
\end{table}

The ramp function was chosen as optimal in this case as it closely follows the independent depth mode at most depths, and uses only two parameters. Note that the number of objects in each depth bin is not uniformly distributed, as $88\%$ of the object observations are at depths $\leq 2(x)$, with about $4\%$ of observations at depths $\geq 4(x)$, and the remaining $8\%$ lie in the linearly-varying portion of the ramp function.
\begin{linenomath*}
\begin{equation}
D(z) =  \begin{cases} 
      D_{1}; & z \leq z_{1} \\
      D_{1} + \frac{D_{2}-D_{1}}{z_{2} - z_{1}}(z-z_{1}); & z_{1} \leq z \leq z_{2} \\
      D_{2}; & z > z_{2}
   \end{cases}
\label{eq:depth}
\end{equation}
\end{linenomath*}
The ramp model in depth is used in both Model 1 and 2 and given by the piece-wise function in Equation \ref{eq:depth} where $z_{1}$ and $z_{2}$ were set to the $2(x)$ and $4(x)$ values in Figure \ref{figure:depthvariationplot}, respectively. A possible physical explanation for the ramp function is described in Section \ref{section:discussion}.

\subsection{Brightness Model Overview}

Three least squares models were developed. The first model, termed `Model 0' (Equation \ref{eq:model0}), was developed to determine the inherent variability of the brightness statistic, which gives context to other models' residuals. Model 0 is designed to provide insight into the best possible performance achievable by other reasonable models of brightness. To do so, Model 0 contains an independent variable for every unique instance of the 449 objects for each of the 5 arrays. In the context of this dataset, Model 0 acts as an upper bound on model performance, as inherent variability and imperfections in the dataset are known to exist.

`Model 1' (Equation \ref{eq:model1}) utilizes the dependent, categorical object name variable, whereas `Model 2' (Equation \ref{eq:model2}) is limited to the objects' physical attributes such as: shape, size, and metallic content. Model 1 has more parameters, and as a result, gives a finer level of granularity to the data. On the contrary, Model 2 is less granular, but aptly fit for prediction of new objects not already present in this dataset. To predict the brightness statistic for a new unobserved object, Model 1 requires speculation of which existing object name is most similar the new object. Model 2, on the other hand, requires only the object's physical attributes, and is therefore capable of predicting any unseen object. 

Among the three models, Model 0 has the most parameters and best fits the data, intentionally, while Model 2 has the fewest parameters and the poorest fit to the data. An overview of the models' variables and goodness-of-fit can be found in Appendix \ref{appendix:modeloverview}. Additionally, the condition number for each model is provided to illustrate the linear independence that these models' variables have, which is part of the reason sum-zero constraints were placed on categorical variables. Each parameter includes a $P > |t|$ statistic (\textit{p}-value). Because of the large number of observations and comparatively small number of parameters, the \textit{p}-values indicate the probability of the null hypothesis (parameter is not important) standing true. The uncertainty in the parameters is also given, specifically, at a 95\% confidence interval. Finally, the number of unique objects and observations that contributed to each variable is also provided, which is particularly relevant for categorical variables.

\subsection{Brightness Model 0: Inherent Variability}
\begin{linenomath*}
\begin{equation}
SLI_{\text{Model\ 0}} = A_{\text{objectID,array}}
\label{eq:model0}
\end{equation}
\end{linenomath*}
Nominally, the same object observed by the same array and configuration should produce the same brightness statistic. However, this is not the case, as a variety of factors contribute to experimental error, leading to some unknown variation in brightness, which we term here the ``inherent variability''. The design of Model 0 is to provide insight into the inherent variability of the dataset.

To determine the inherent variability, a unique model parameter was created for each object identifier and array combination (denoted $A_{\text{objectID,array}}$ in Equation \ref{eq:model0}). In other words, 449 unique object identifiers and 5 arrays gives 2,245 unique variables. However, each array did not observe every object, so 2,101 unique variables arise. Since this synthesized variable is categorical, the coefficients were constrained to a sum of zero and dummy encoded, which would lead to 2,100 parameters, however, the additional constant term brought the total parameters back to 2,101. 

\begin{figure}
	\centering
	\includegraphics[width=3.4in]{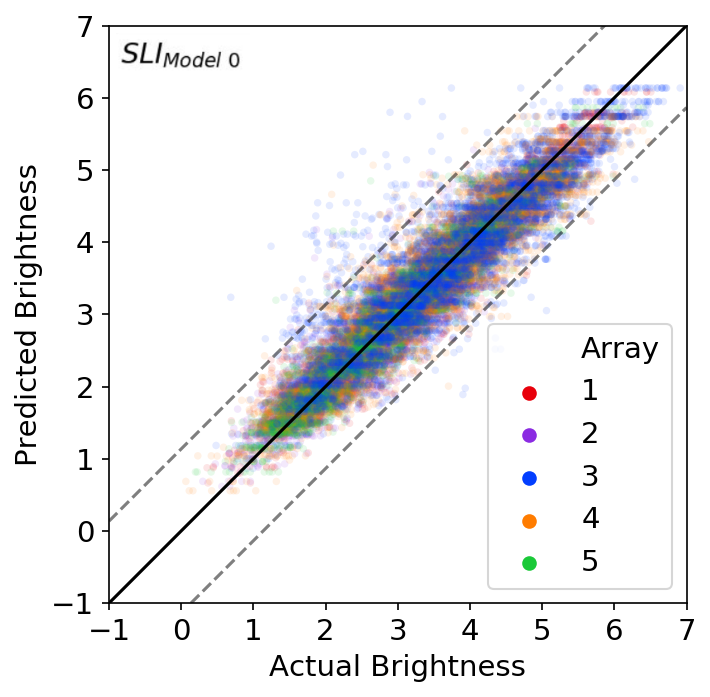}
	\caption{Scatter plot of Model 0 predicted and observed brightness values, colored by array. Where the dashed black lines indicate the predicted and observed brightness values are within $3\sigma_0\approx1.1$ SLI units.}
	\label{figure:Model0scatterplot}
\end{figure}

Fitting all 25,643 observations to 2,101 free variables gives an $R^2_{\text{adj.}}$ of $0.88$. The root-mean-squared (RMS) of the residuals is $0.38$ SLI units, which corresponds to this inherent variability, hereafter denoted by $\sigma_0$. A plot of the brightness values predicted by this model, versus their observed brightness, is given in Figure \ref{figure:Model0scatterplot}. $98.9\%$ of predictions fall within $3\sigma_0$ of their observed brightness, which is slightly less than the $99.7\%$ that would be expected for a normal distribution.

In Figure \ref{figure:Model0scatterplot}, notice some observations' actual brightness values are dimmer than predicted (\textit{i.e.,} the points above the upper dashed line). Analyzing the residuals from this model, it was determined that those data points are primarily associated with GPS uncertainty. In other words, the uncertainty radius was not large enough to capture the peak. However, since there is a unique coefficient for each object identifier and array, for an observation to fall above that dashed line means a majority of other observations of the object \textit{did} fall within the uncertainty radius. The origin of these GPS uncertainties was traced to hardware limitations, so the only way to compensate for the uncertainty would be choosing a larger radius.

However, some points fall below the lower dashed line, which corresponds to brighter observations than predicted. For these points, the object was often emplaced near clutter, meaning that the uncertainty radius captured both the object itself and nearby clutter, which was occasionally brighter than the object itself. Figure \ref{figure:1}d illustrates this case, where the maximum brightness is coming from the clutter along the top edge of the uncertainty radius, instead of from the center where the object can be seen, likely due to imperfect surface removal. More sophisticated image processing algorithms could be employed to compensate, but again, this was neglected in favor of simplicity. A smaller uncertainty radius could improve the fit for these brighter-than-expected observations, but is in conflict to the dimmer-than-expected observations, which would benefit from a larger radius. Thus, in the context of the large population size considered, the $25$cm uncertainty radius establishes a reasonable balance between GPS uncertainty and nearby clutter.

Model 0 achieves its goal of providing insight into the inherent variability present in this specific dataset. However, given the magnitude of information and independent variables present (many objects observed by many arrays), Model 0's insight might reasonably be considered as depictive of the inherent variability in GPR systems as applied to EHD and beyond.

\subsection{Brightness Model 1: Name-Based Model}
\begin{linenomath*}
\begin{equation}
SLI_{\text{Model\ 1}} = B_{\text{object\ name}} + C_{\text{array}} + D(z)
\label{eq:model1}
\end{equation}
\end{linenomath*}
Model 1 is given in Equation \ref{eq:model1} and uses three independent variables: array, object name, and object depth. $C_{\text{array}}$ is categorical across 5 different arrays meaning 4 free parameters with a zero-sum constraint, specifically given in Equation \ref{eq:zsum} and $B_{\text{object\ name}}$ is also categorical with 66 different values, implying 65 free parameters.
\begin{equation}
\sum_{\text{object name}} B_{\text{object name}} = \sum_{\text{array}} C_{\text{array}} = 0
\label{eq:zsum}
\end{equation}

$B_{\text{object\ name}}$ is a granular variable, which provides the least squares regression with an expanded capability to control for different objects and in some cases different poses. For example, the object name variable specifies different pose-dependent categorical levels for certain projectiles. Interestingly, in a number of cases, the same object in different poses was found to be both statistically significant and different. For depth, the ramp function from Section \ref{depthvariation} is used. Rather than constraining the depth function to be sum-zero, the intercept (constant) is incorporated into the depth function, leading to two free parameters in depth. Together, there are 71 parameters in this model.

\begin{figure}
	\centering
	\includegraphics[width=3.4in]{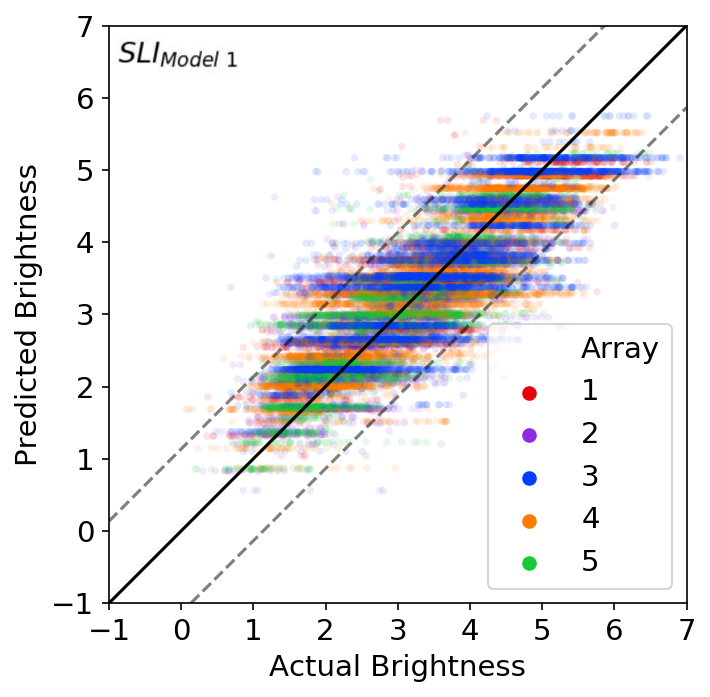}
	\caption{Model 1 predicted and observed brightness, colored by array. Where the dashed black lines indicate the points are within $3\sigma_0\approx1.1$ SLI units.}
	\label{figure:Model1scatterplot}
\end{figure}

Model 1 achieves an $R^2_{\text{adj.}}$ of $0.70$, and the residual RMS is $0.62$ SLI units, which corresponds to about a 60\% increase in the residual RMS of Model 0. Another plot of predicted vs actual brightness can be found in Figure \ref{figure:Model1scatterplot}. For this model, $93\%$ of the predicted brightness values are within $3\sigma_0$ of the actual values. Worthy of note is the difference of $R^2_{\text{adj.}}$ values between Model 0 and Model 1 in the context of size. While Model 1 is ``less fit" than Model 0, its performance is remarkable when considering 71 vs. 2,101 free parameters. Effectively this means a 97\% reduction in model capacity with only a 8\% loss in $R^2_{\text{adj.}}$ fitness from the maximum possible. 

Model 1's full parameters can be found in Appendix \ref{appendix:Model1}, and further discussion of their interpretation can be found in Section \ref{section:discussion}. Note that this appendix only provides object name coefficients if that object name has at least 5 unique objects and 100 total observations available in the dataset.

\subsection{Brightness Model 2: Property-Based Model}
\begin{linenomath*}
\begin{equation}
\begin{aligned}
SLI_{\text{Model\ 2}} = C_{\text{array}} + D(z) +E_{\text{object\ type}} + F_{\text{area}} * Area
\end{aligned}
\label{eq:model2}
\end{equation}
\end{linenomath*}
\begin{figure}
	\centering
	\includegraphics[width=3.4in]{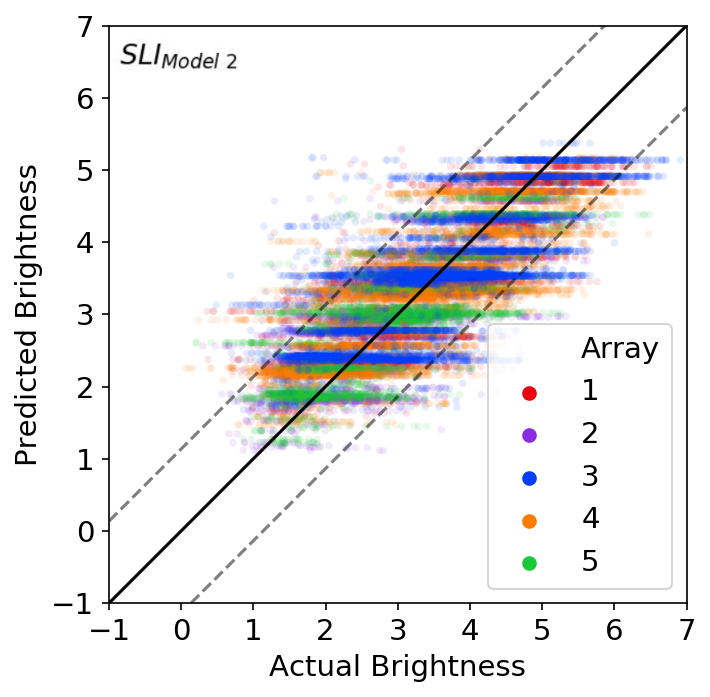}
	\caption{Model 2 predicted and observed brightness, colored by array. As before, the dashed black lines denote points are within $3\sigma_0\approx1.1$ SLI units.}
	\label{figure:Model2scatterplot}
\end{figure}

Model 2, unlike Model 1, is not based on the object's name, but rather based solely on the object's physical attributes. (Equation \ref{eq:model2}) The following variables are used in Model 2: object type, bounding box area, as well as depth and array. Depth utilizes the same ramp function defined for in Model 1 and bounding box area ($F_{\text{area}}$) is a numeric variable whose coefficient is a simple proportionality constant $Area$. Finally, $E_{\text{object\ type}}$ is a simple 10-level categorization of the greater family of ``object types" a particular object belongs to. See Section \ref{objectdata} for further details on these variables. Taken together, there are 17 model parameters in Model 2, approximately a 76\% reduction in model size from Model 1.

For this model, $R^2_{\text{adj.}}$ is $0.62$, and the residual RMS is $0.71$ SLI units, which is a $13\%$ increase in residual variance as compared to Model 1. $90\%$ of predicted brightness values fall within $3\sigma_0$ of their actual brightness. A plot of the predicted and actual brightness can be found in Figure \ref{figure:Model2scatterplot}.

The model parameters can be found in Appendix \ref{appendix:Model2}, and further discussion of their interpretation can be found in Section \ref{section:discussion}.

\subsection{Simulation Experiments}

In order to test models of object brightness, a simulation scheme was devised as follows:
\begin{enumerate}
	\item Randomly split the 25,643 observations into two parts, training data and test data, each consisting of 50\% of the data.
	\item Fit models using the training portion of the observation data.
	\item Predict brightness values in the test sample.
	\item Record the proportion of predictions within $\pm3\sigma_0\approx1.14$ SLI units of the actual visibility values. (absolute error)
	\item Repeat steps (1)-(4) over 1,000 iterations
\end{enumerate}

The inception of this design was inspired by several factors. The terse nature of visibility is subject to some unknown level of inaccuracy stemming from both environmental and systemic factors, making it difficult to identify and eliminate errors. Additionally, a prediction within $\pm3\sigma_0$ SLI units was considered to be the most important measure of a model's success. A model built using half the data and routinely predicting brightness with an absolute error $\leq3\sigma_0$ was considered sufficient generalization.

\begin{figure}
	\centering
	\includegraphics[width=3.4in]{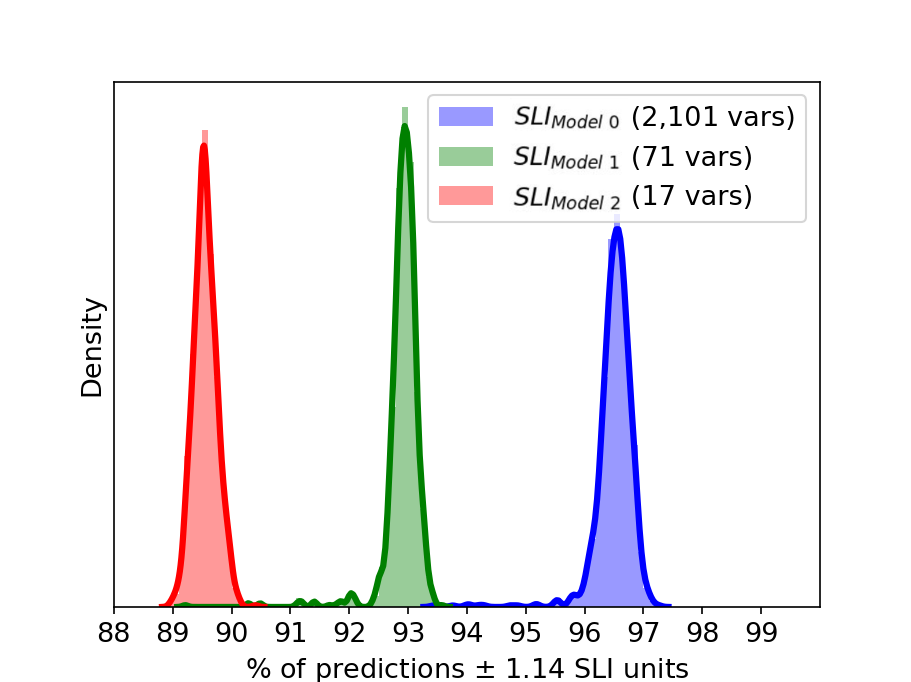}
	\caption{Percent of model predictions within $3\sigma_0$, with the models trained on half of the observation data, then used to predict the other half's brightness values.}
	\label{figure:simulations}
\end{figure}

By randomly splitting the data in half for training and test sets, there was no guarantee that the models would be well-conditioned. For example, an object name with few observations could randomly all lie within the test set, meaning the training data is unable to predict the coefficient associated with that object name (and the coefficient is thus treated as zero). In the 1,000 iterations of this experiment, Model 1 had 37 iterations with poor conditioning. Model 0 was poorly conditioned for all 1,000 iterations, and Model 2 was well-conditioned for all 1,000 iterations. 

In Figure \ref{figure:simulations}, Model 2, with just 17 parameters, predicted $89.7\%$ of the data to within $\pm3\sigma_0$. It has the advantage of generalizing to other objects not seen before; just the object type and area are needed, along with the object's depth and which array was used. While Model 1 had better predictive power $92.9\%$, it relies on object name. Thus, it cannot easily predict a new, unobserved object. Model 1 does possess the ability to predict a new object's brightness, just at a new depth, with a different array, or by associating a new object with one of 66 object names. Model 0 had the best predictions ($96.6\%$), but its unreasonable number of model parameters (2,101) makes it effectively useless for predicting object brightness, as it requires the training data to have that combination of array and object already available. For this reason, Model 0 serves best as a baseline for measuring the inherent variability in GPR applications.

Additional results were tabulated using a threshold of $2\sigma_0$, where Model 0 captured $91.2\%$ of the test data, Model 1 had $79.2\%$, and Model 2 covered $73.9\%$.

\subsection{Variable Significance}

In addition to the independent variables used in Models 1 and 2, a number of other independent variables were considered, including array height, vehicle speed, and location (\textit{i.e.,} testing facility) -- see Section \ref{rundata} -- in addition to the number of days between object emplacement and observation (for the possibility of degradation over time).

All four of these variables were found to be statistically significant, in that they increased the $R^2_{\text{adj.}}$ and had nearly-zero $p$-values, but were found to be \textit{practically} insignificant, in that their effect on predicted brightness were negligible compared to $\sigma_0$. Incorporating all four variables (which constituted an additional five model parameters) led to an increase in $R^2_{\text{adj.}}$ of just 0.0065 in Model 1, and 0.0079 in Model 2, while the residual RMS SLI values only decreased by 0.007 and 0.008 SLI units in Model 1 and Model 2, respectively.

Of the independent variables that were used in Models 1 and 2, all were more important than the variables denoted above which were omitted. In Model 1, object name constituted the largest fraction of the explained variance, followed by depth, and then array. In Model 2, an object's general type explained the largest fraction of observed variance, followed by depth, bounding box area, and array, in order. In both models, array having the least amount of explanatory power might be due to the SLI image standardization removing the greatest amount of variability between arrays.

\subsection{Effects of Incomplete Information}

A study of buried object brightness based on experimental data is naturally limited by dataset scope and completeness. Therefore, while the dataset size was sufficient for modeling, incomplete information and a lack of diversity in some object attributes played a role in guiding our efforts. 

One notable property which lacks sufficient diversity in our dataset is the relative soil conductivity. The soil type an object is buried in has previously been shown to have a significant impact on detection performance for certain objects. \cite{yuksel2019transfer,colwell2015improving} While we did consider an object's location to evaluate this effect, the two locations represented in the data set have similar soil types (primarily sandy loam) and we believe relative soil conductivity would have a significant effect on object brightness upon collecting data in more diverse locations.

We otherwise consider the specific lack of information and sampling bias in this data set to be adequate for object brightness modeling in EHD applications, but note that biases may vary across the larger field of GPR.

Specifically, in GPR's application to EHD, it is common for certain object properties to be unknown, particularly when considering improvised objects. As a result, some object properties were found to be significant for brightness estimation, but excluded from the final models in an effort to improve model generalization.

\section{Object Visibility}
\label{section:ObjectVisibility}
Buried object brightness models create a means for quantitatively comparing buried objects, but do not directly answer whether or not an individual observation of an object appears visible. As previously described, an object may not appear visible in individual observations due to GPS error or the inherent variability of GPR data, as well as some objects lacking sufficient contrast with the background to be visible. Here, visibility refers to whether or not the standardized brightness value (\textit{i.e.,} contrast) associated with that observation is truly representative of an object, as opposed to background clutter.

While a cut-off brightness value for visibility could be nominally chosen, the threshold might vary across use cases, arrays, and individual observations. In order to close the gap between a buried object's brightness and whether or not the object is visible in an individual observation, we formulated and implemented the concept of probability of visible or $P(visible)$.

$P(visible)$ is an expression of the probability an object is visible based on the max brightness observed and the radar array used to observe it. Our goal is simple: Given a brightness value from a known object's ground truth quadrilateral, what is the probability the brightness value represents an object?

\begin{figure}[h!]
	\centering
	\includegraphics[width=3.4in]{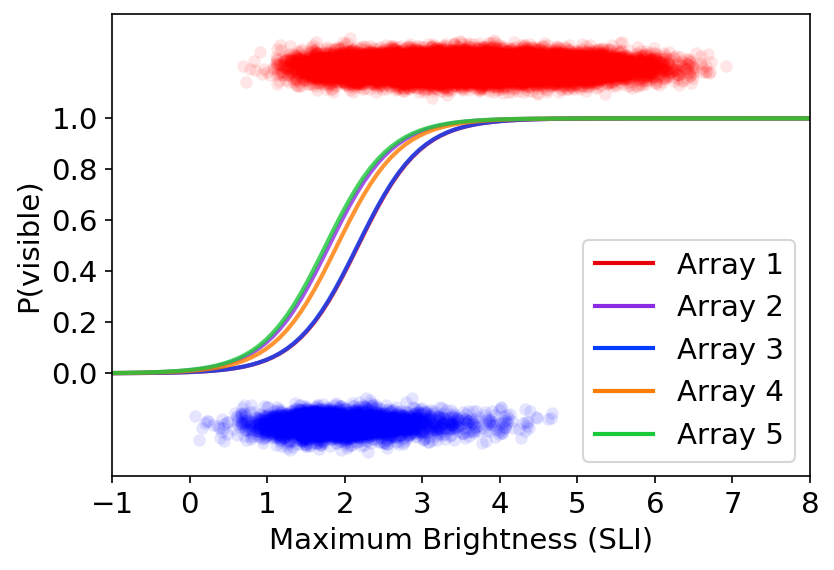}
	\caption{Logistic model of $P(visible)$, split by array. The points in the upper region indicate brightness values of visible objects (red), and the lower points indicate brightness values of background (blue).}
	\label{figure:pvisible}
\end{figure}
\subsection{Dataset Design}
To model $P(visible)$, the dataset was split into visible (21,328 red points) and background (4,315 blue points) observations, see Figure \ref{figure:pvisible}. This split was done manually, where images of each of the 449 objects were assessed on their visibility to the human eye. While there is some subjectivity in the determination of an object's visibility in an image, the majority of observations are fairly unambiguous in their visibility. For example, in Figure \ref{figure:1}, all three objects (recall that panels \textit{a}) and \textit{b}) are the same object) are labeled visible. Panel \textit{d}) is a borderline case, and it was only after considering other observations of this object that it was deemed visible. Additionally, a correlation of this subjective visibility and our final $P(visible)$ model was performed and showed no notable discrepancies.

The ``background" observations were crafted carefully. Only objects which never appeared visible across all radar arrays and all observations were taken to belong to the background class. Interestingly, this formulation makes predicting $P(visible)$ more difficult than comparing brightness values from known objects to, for example, brightness values from locations where objects are not emplaced. In that, the background class isn't truly sampled from a distribution of random noise. Instead, each background observation comes from a location where objects are known to exist, just not visibly. As such, the background observations contain a very specific bias, disturbed earth. If the max brightness values in the dataset are biased higher because of a disturbed earth effect, the background samples in this case will be representative of that bias. In Figure \ref{figure:pvisible}, most visible objects have higher values than backgrounds on average, but there is no single threshold that perfectly splits the visible objects from background.
\subsection{Model Design}
In order to model $P(visible)$ we considered logistic regression and CART classification trees \cite{doi:10.1080/00401706.2020.1744905}. Ultimately we elected to use a logistic regression model for $P(visible)$, as a careful reading of the tree, despite being provided array and max brightness, revealed a simple binary threshold above and below a max brightness of 1.49. Instead, the logistic model was used. It had 6 total parameters. A $b$ value for each array (which gives the curves their horizontal position), and one scaling parameter for the brightness statistic, $m$ (which gives the curves their horizontal width), and is independent of array. The $P(visible)$ model equation is given by:
\begin{equation}
	\label{eqn:pvis}
	P(\text{visible}|\text{SLI}) = \frac{1}{1+\exp\left(-\left(m~\text{SLI}+b\right)\right)}
\end{equation}

The logistic model is shown in Figure \ref{figure:pvisible}.

Table \ref{table:pvisparams} shows the probability of an object appearing visible, given its maximum brightness value shown in the table. In other words, for an object with a maximum brightness of approximately 2.4, there is a 95\% chance it is a visible object for Array 3, as opposed to simply being a measurement of the background.
\begin{table}[!h]
\begin{center}
	\begin{tabular}{cccccc}
		\toprule
		{} &     $m$ &       $b$ &    55\% &    75\% &    95\% \\
		\midrule
		Array 1  &  2.42 & -3.248 &  1.425 &  1.797 &  2.560 \\
		Array 2  &  " & -2.440 &  1.091 &  1.463 &  2.226 \\
		Array 3   &  " & -2.790 &  1.236 &  1.607 &  2.370 \\
		Array 4 &  " & -2.647 &  1.177 &  1.548 &  2.311 \\
		Array 5   &  " & -2.325 &  1.044 &  1.415 &  2.178 \\
		\bottomrule
	\end{tabular}
\end{center}
	\caption{$P(visible)$ model parameters by array in Fig. \ref{figure:pvisible}}
	\label{table:pvisparams}
\end{table}

\subsection{Sensitivity vs. Specificity}

Our formulation of $P(visible)$ as a logistic regression model provides an even more granular pathway to analysis. By labeling every observation with a probability the object appears visible, a trade off arises. That is, determining where to threshold $P(visible)$ to capture objects and eliminate background clutter.

\begin{figure}[h!]
	\centering
	\includegraphics[width=3.4in]{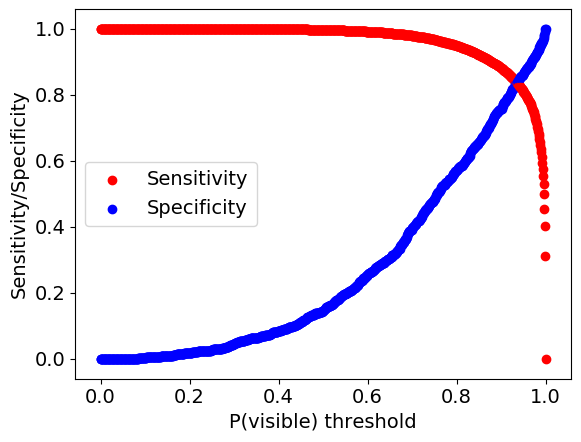}
	\caption{Sensitivity and specificity for $P(visible)$ over all observations and arrays.}
	\label{figure:sense}
\end{figure}

Figure \ref{figure:sense} demonstrates the advantages and disadvantages of choosing $P(visible)$ thresholds in the 0-1 range of probabilities. Sensitivity refers to the probability of correctly identifying the presence of a visible object in an individual observation, $P(present|visible\,object)$ and is often known as \textit{power} in hypothesis testing. On the other hand, specificity is the probability of correctly identifying the absence of a visible object, $P(absent|invisible\,object)$. 

While the correct $P(visible)$ threshold is not an objective number, it provides the qualitative information necessary to choose a value appropriately. For example, selecting a cut-off value of $P(visible) = 0.65$ (x-axis) corresponds to a sensitivity of 0.987 and a specificity of 0.305. Thus, if all values with a $P(visible) \geq 0.65$ are considered, then observations of a visible object will be correctly identified 98.7\% of the time. Conversely, for observations with $P(visible) \leq 0.65$, the absence of a visible object is correctly predicted 30.5\% of the time. Table \ref{table:cutoffs} summarizes some key cut-offs.

\begin{table}[!h]
	\begin{center}
		\begin{tabular}{ccc}
			\toprule
			\textbf{cut-off} & \textbf{sensitivity} & \textbf{specificity} \\
			$P(visible)$ & $P(present|visible)$ & $P(absent|not\,visible)$ \\
			\midrule
			0.55  &  0.995 &  0.198   \\
			0.65   & 0.987 &  0.305  \\
			0.75   & 0.968 &  0.485  \\
			0.85 &  0.922 &   0.658  \\
			0.95  & 0.810 &   0.950  \\
			\bottomrule
		\end{tabular}
	\end{center}
	\caption{Sensitivity and Specificity for key $P(visible)$ thresholds across all arrays.}
	\label{table:cutoffs}
\end{table}

The applications of $P(visible)$ models span from validating automatic detection algorithm training data to qualifying algorithmic performance. A standalone $P(visible)$ model retains the ability to act as a baseline detection algorithm. While we are not recommending its use for such a task, applying a $P(visible)$ regression model to radar data can provide a lower-bound on how well a more sophisticated algorithm should perform. Additionally, when analyzing shortcomings of a sophisticated detection algorithm, $P(visible)$ probabilities can aid in quantitatively indicating whether a detection algorithm failed (the object has a high probability of being visible) or the radar hardware failed (the object has a low probability of being visible). This represents a portion of our intention for $P(visible)$ modeling. GPR detection algorithms can fairly report their performance at different $P(visible)$ thresholds to separate system-level performance (hardware + algorithm, $\lim_{P(visible) \to 0}$) from pure algorithmic performance. The secondary intention for $P(visible)$ is as a mechanism to remove background observations from training data (such as for a convolutional neural network). Modern machine learning approaches benefit from training dataset cleanliness (\textit{i.e.,} no background observations labeled as targets) and using a $P(visible)$ threshold at, for example, 0.55 would successfully allow for filtering out approx. 20\% of mislabeled positive (visible object) examples from a training set, while retaining over 99\% of true positive examples.

\section{Discussion}
\label{section:discussion}
The insights gained from both forms of models intuitively align with well-known principles of GPR and nuances of GPR as applied to EHD. The information provided by our models serves to provide qualitatively-based claims with regard to the performance of GPR systems, the objects those systems observe, and insight into the greater landscape of GPR applications. By analyzing the interactions between a system, its environment, and the objects buried, our models parse out which independent variables effect a GPR system and how much of an effect they have. Several important takeaways arise, the most significant of which are detailed below.
\subsection{Comparison of Model Parameters}
\subsubsection{Array}
Array is found in both models and the same trends are found in each (see Appendices \ref{appendix:Model1}, \ref{appendix:Model2}). The conclusion that arises and is agreed upon by both models is the statistical superiority of Array 3 to all other arrays, followed by Array 1, Array 4, Array 5, and finally Array 2. An interesting takeaway from the 95\% confidence intervals around the array coefficients for both models is that the brightness modeling shows a statistically significant difference between each array and the others.

\subsubsection{Shape}
In Model 2, shape was considered as a variable. The inclusion of shape as a 6-level categorical variable only gave a 0.0073 improvement to Model 2's $R^2_{\text{adj.}}$ fitness and therefore it was excluded from Model 2's final design. However, there are some interesting takeaways from the categorical coefficients when shape is included. 

It is found that square objects are noticeably brighter than circular objects. One plausible explanation for that is that, for equivalent bounding box area, a square is simply larger than a circle, by approximately 27\%. However, another explanation is that corners are more reflective than edges, which are more reflective than faces, due to the effects of diffraction \cite{1138719}. Furthermore, many objects have irregular shapes, which complicates their categorization.

Ultimately, the inclusion of object shape made Model 2 poorly conditioned. We believe this phenomena occurs because our definition of object type categories causes objects of certain shapes to fall in the same object type categories.

\subsubsection{Depth}
Since the ramp function was used in both Model 1 and Model 2, the best fit coefficients can be compared. The difference in brightness between shallow and deep portions of the ramp function are comparable between the two models, approximately 1.6 SLI units. However, their overall offset from 0 is different, by about 0.6 SLI units. Model 1's other parameters are all categorical, and therefore are constrained to be mean zero. Model 2's other parameters are also mostly categorical, except for the bounding box area variable, which is simply a proportionality constant. Since the average object's area is $0.077m^2$, this equates to a typical offset of 0.45 SLI units, leaving just 0.15 SLI units left unaccounted for between the two depth models. Therefore, due to the formulation of these two models, the relative effects of depth are approximately the same, but the absolute depth coefficients are not as meaningful to compare directly.

\subsubsection{Anti-Tank vs Anti-Personnel Objects}

It's not uncommon for GPR detection systems in EHD to be applied or evaluated against anti-tank (AT) or anti-personnel (AP) objects exclusively \cite{1356065, 1642579, 4779004, 5650741, Song_2019, 8465980, 10.1117/12.484177}. Specifically, in \cite{10.1117/12.2176250}, the general notion that AP objects are more difficult to detect is clearly shown, as AP mines present more difficultly due to smaller size and less metallic content \cite{andrews_ralston_tuley_1999}.

However, associating all objects with either family is not straightforward. Using the well-defined AT/AP classification for brightness modeling eliminated 102 objects and approx. 5,000 observations from the data set. Alternatively, adding an additional unknown category, while statistically significant, provided no benefit to the model's $R^2_{\text{adj.}}$ fit. Worth noting is the agreement of the regression fit with the notion that AP objects are more difficult to detect than AT objects, \cite{ 10.1117/12.2176250} with the coefficients indicating AP objects are not as bright as AT objects.

Our hypothesis is the AT/AP categorization as an explanatory variable has negligible impact on model fitness, because metallic content and object size were both already utilized as input in Model 2.

\subsection{Depth Dependence with Clutter and Noise}
\label{subsection:depthclutternoise}
The ramp function for depth variation in maximum brightness was chosen primarily based on the similarity to the independent depth function illustrated in Fig. \ref{depthvariation}, and its corresponding goodness of fit compared to other two-parameter functions, such as the linear fit also shown in \ref{depthvariation}. However, some physical explanation can be attributed to this ramp function.

The standardization process, detailed in \cite{SLIpaper}, compares the object's image intensity to the background image intensity at a comparable depth as the object. The image intensity at the object can be considered to be $I_0 \exp(-2\alpha z) \sigma_{\text{object}}$, where $I_0$ is the incident intensity on the object in the absence of attenuation, $\alpha$ is the attenuation coefficient, and $\sigma_{\text{object}}$ is the object's radar cross section. In a clutter-limited environment, the clutter's image intensity would be $I_0 \exp(-2\alpha z) \sigma_{\text{clutter}}$, however, in a noise-limited environment, the image intensity would be $I_{\text{noise}}$. The standardized logarithmic intensity would be the logarithm of the ratio between the object image intensity and the background image intensity. Therefore, in a clutter-limited environment, the SLI value would be proportional to $\log\left(\sigma_{\text{object}}/\sigma_{\text{clutter}}\right)$, whereas in a noise-limited environment, the SLI value would be proportional to $\log\left(I_0 \sigma_{\text{object}}/I_{\text{noise}}\right) - 2\alpha z$. 

In other words, in a clutter-limited environment, the SLI value is independent of depth, whereas in a noise-limited environment, the SLI value would be linearly decreasing in depth. Thus, the shallow, constant portion of the ramp function physically corresponds to clutter-limited, and the linearly decreasing portion of the ramp function physically corresponds to noise-limited, with the transition occurring at the shallow inflection point, $z_1$. The object's brightness would continue decreasing without bound, except for the fact that, eventually, the object `signal' would be drowned out by the noise, leading to brightness values that are statistically drawn from a normal distribution. Thus, there is a practical lower-bound for brightness values, which could be considered part of the deep, constant portion of the ramp (\textit{i.e.,} $z > z_2$).

\subsection{Brightness Model 1}

Model 1 shows an overwhelming agreement with and provides quantitative support for several nominally accepted principles in EHD. While the object names are obfuscated for sensitivity, different objects considered to be similar in nominal difficulty group together in either the upper, middle, or lower quadrants of coefficients. The same can be said of the radar arrays used to collect observations. The arrays nominally considered to be better have positive coefficients, while the arrays considered to be lower performing have negative coefficients.

Consider the following cases: a shallow low-metallic mine, and a deep metallic mine. A shallow ($\leq z_1$) low-metal mine seen by the Array 3 would have a predicted value of 4.02 (the depth, object name and array give values of $+3.248$, $+0.502$ and $+0.270$, respectively). A deep ($z>z_2$) metal mine seen by Array 3 would have a predicted value of 4.01 (the depth, object name and array give values of $+2.740$, $+0.998$, and $+0.270$). In other words, these two different object observations would appropriately be predicted to have almost equivalent brightness.

One concern is Model 1's usage of 66 object name categories, this specificity could lead to over-fitness. However, in the simulation scheme which only provides 50\% of the dataset for the model to train on, this concern is mitigated as Model 1 consistently shows its strong performance generalizes and is rarely ill-conditioned.

\subsection{Brightness Model 2}

Using Model 2, it is possible to compare the visibility differences between, for example, a large non-metallic, improvised object, and a small non-metallic, improvised object. For the same shallow depth, and the same array (\textit{e.g.,} Array 1), a 15 square inch object would have a brightness of 3.29 (the array, object type, depth, and area effects are, respectively: $+0.189$, $-0.064$, $+2.965$, and $4.007\times0.0507m^2=+0.2032$). On the other hand, a 7 square inch object would have a brightness of 3.10 (the array, object type, depth, and area effects are, respectively: $+0.189$, $-0.064$, $+2.965$, and $4.007\times0.0032m^2=+0.0128$). From these coefficients, we can understand the relative importance of object size and depth. Varying depth has a stronger impact (larger coefficient), but doubling the size of an object still effects the object's brightness. 

The insights Model 2 is capable of providing are a unique utility. Instead of needing to know \textit{what} object is in the ground in order to estimate brightness, Model 2 asks only to know the properties of the object. While its $R^2_{\text{adj.}}$ fitness doesn't exhibit ideal performance, we believe more data and object information can aid improvements. This is expected, as the scope of the problem Model 2 is asked to solve is much more general than both Model 1 and especially Model 0. Conservatively, Model 2 is tasked with solving for the causal relationship between GPR systems and objects within the context of EHD and multi-static radar systems. While its exact parameters may need adjustment for different applications and systems, the intuition behind Model 2's development is easily transferred across the landscape of GPR.

\subsection{Visibility Model}

Applying a logistic regression to brightness values is straightforward, but serves to quantify what is nominally apparent via visual inspection of the tomographic data. Despite the significant differences in architecture from our brightness modeling, $P(visible)$ reveals important details about each array's signal and noise. When considered together, we believe $P(visible)$ is capable of indicating more general information about GPR systems, at least within the context of multi-static systems applied to EHD. In Figure \ref{figure:sense} and Table \ref{table:cutoffs}, we examine the overarching takeaways from all array $P(visible)$ models considered together. 

While $P(visible)$ is a poor detection algorithm itself, it provides a capability for failure analysis, dataset cleaning, and detection algorithm base-lining. Additionally, we believe $P(visible)$ solves a major pitfall of current GPR system analysis. With Receiver Operating Characteristic (ROC) curves as the primary mode of GPR analysis, especially in EHD, a conundrum emerges. Which objects do and do not belong in a performance analysis? Of course, the answer is application and mission specific, but, the exclusion and inclusion of objects as valid grading criterion remains subjective. $P(visible)$ offers a quantitative way to address this question in the field of EHD and the greater landscape of GPR. While we do not eradicate all subjectivity, $P(visible)$ makes a quantitative claim about which objects demand inclusion in performance analysis. If for example, one was to choose objects ad hoc to include and exclude in performance analysis, a $P(visible)$ model can be used to identify any excluded objects which appear at least as visible as those included in the performance analysis, thereby quantitatively nominating a subset of the excluded objects as necessary to include in analysis.

\section{Conclusions and Future Work}
\label{section:Conclusions}

Buried object brightness was conceptualized and modeled to provide a more robust means for GPR system performance characterization. Both forms of our final models (Model 1 and Model 2) showed sufficient accuracy in predicting brightness when provided with half the original data set from five different radar arrays, providing evidence they properly uncover important information applicable across GPR systems, objects, and applications.

Through this modeling effort, a variety of conclusions can be drawn, including: the relative brightness of various buried objects, which arrays create brighter objects, the effects of depth, shape, and metallic content on brightness, the natural variability in the brightness metric, the apparent insensitivity of object brightness to array height and speed, and the visibility of objects based on their observed brightness.

While this model used GPR images of buried explosive hazards, the model machinery may, in principle, be applied in other domains. This could include other GPR domains, such as tunnel detection, petroleum or mineral exploration, or bridge/roadway inspection, or even domains outside of GPR, such as other radar applications, or sonar or x-ray tomography. Additionally, within GPR for EHD applications, signal scans can in principle be used instead of images, particularly if an appropriate standardization method can be applied.

The potential use cases for a model of object brightness and visibility are numerous. One example is a metric by which different GPR systems can be compared to one another, without requiring nearly identical test objects, since with enough data and object diversity, the varying effects of the objects can be controlled for, and statistical claims can be made about the relative performance of different GPR systems. 

Another use case for these models is in the creation of performance metrics, such as ROC curves. Deciding which objects should be included or excluded from a ROC curve is partially guided by the goals of the detection system, but may also be partially guided by anecdotal evidence for which objects are or are not typically visible. This study provides a framework by which object brightness and visibility can be estimated, and can provide quantitative recommendations for object inclusion or exclusion. For example, consider the dimmest object deemed to be appropriate to include in the performance characterization -- all objects brighter than this \textit{should} be included in the performance characterization, unless detecting those objects is not considered germane to the GPR system and application under consideration.

Our $P(visible)$ models also provide a useful capability, in that, they provide an automated, quantitative way to separate algorithmic performance from system performance. In doing so, $P(visible)$ can also be used to clean datasets for more sophisticated classifiers and act as a baseline classifier itself. In sum, the combination of our brightness and visibility models arm GPR research to rapidly identify causes of failure, objectively grade algorithms, all while providing better understanding and potentially improved performance.

\subsection{Future Work}
In future work, we aspire to improve Model 2, populate missing information in the data set, add new radar arrays, and mitigate multicollinearity in the brightness and visibility models. Our study was limited by missing information and therefore subject to potential bias in the paths we chose to pursue. While this lack of information is prevalent in GPR for EHD, other applications of GPR would benefit greatly from an accurate visibility model where the name of an object/anomaly is not required as input. While the nature of the five radar arrays considered varied significantly, an increased number of arrays/systems would inherently make any visibility model based on experimental data more robust. Finally, because of the limited field of object properties and array configuration variables available, a multicollinearity was evident in our brightness models (Appendices \ref{appendix:Model1}, \ref{appendix:Model2}). Simulations fit to half the dataset mitigated concern of this observation, but multicollinearity does impact the theoretical generalization of our models.

In addition to buried object brightness modeling, we presented a $P(visible)$ model to discriminate between observations of an object which are visible and not visible. Using a baseline of background samples, our $P(visible)$ model is an accurate means of discriminating between individual object observations which are subject to error in GPS accuracy and systemic processing errors.

For buried object brightness and visibility modeling, increased samples and sampling diversity in various locations (soil types/conditions) would strengthen the generalization and accuracy of the models. A similar strength would be gained with the addition of observations from new, unseen radar arrays and object types.

\appendices
\onecolumn

\section{OLS Model Overview}
\label{appendix:modeloverview}
\begin{center}
	\vspace*{5mm}
	\begin{tabular}{ccccccc}
		\toprule
		{}		& {} &	Residual	&	{}		& Fraction within*				& Condition	& No. of \\
		Model	& Independent Variables &	RMS		&	$R^2_{\text{adj.}}$	& $\pm3\sigma_0$ ($\pm2\sigma_0$) &
		Number	 & Parameters\\
		\midrule
		0	& Object Identifier$\times$Array &	0.3786 ($\sigma_0$)	&	0.8812 & 98.9\% (95.3\%) & 160.2 & 2,101 \\
		1 & Object Name, Depth, Array &	0.6232	&	0.7036 & 93.1\% (79.3\%) & 121.3  & 71 	 \\
		2 &	Object Type, Area, Depth, Array & 0.7051	&	0.6214 & 89.6\% (73.9\%) & 38.3 & 17 \\
		\bottomrule
	\end{tabular}
\end{center}
	\vspace*{1mm}
* Where $\sigma_0$ refers to 0.3786, the residual RMS for Model 0

\section{Model 1 Regression Parameters}
\label{appendix:Model1}
\begin{center}
	
	\vspace*{5mm}

	\begin{tabular}{cccccc}
		\toprule
		{}  &  Coefficient  &  Uncertainty  &  \# of Unique  &  \# of Unique  &  {}\\
		$C_{\text{array}}$  &  (SLI units)  &  (95\% Conf. Int.)  &  Objects  &  Observations  &  $P > |t|$\\
		\midrule
		Array 3   &  $+0.270$ &  $\pm~0.013$ &  448 &  9,142 &  $<1e-3$ \\
		Array 1  &  $+0.210$ &  $\pm~0.021$ &  427 &  2,532 &  $<1e-3$ \\
		Array 4 &  $+0.039$ &  $\pm~0.014$ &  415 &  6,996 &  $<1e-3$ \\
		Array 5   &  $-0.247$ &  $\pm~0.017$ &  381 &  4,095 &  $<1e-3$ \\
		Array 2  &  $-0.273$ &  $\pm~0.020$ &  430 &  2,878 &  $<1e-3$ \\
		\bottomrule
	\end{tabular}
	
	\vspace*{5mm}
	
	\begin{tabular}{cccccc}
		\toprule
		{}  &  Coefficient  &  Uncertainty  &  \# of Unique  &  \# of Unique  &  {}\\
		$D(z)$ &  (SLI units)  &  (95\% Conf. Int.)  &  Objects  &  Observations  &  $P > |t|$\\
		\midrule
		$D_{1}$  &  $+3.248$ &  $\pm~0.029$ &  422 &  23,977 &  $<1e-3$ \\
		$D_{2}$ &  $+1.621$ &  $\pm~0.046$ &   68 &   1,665 &  $<1e-3$ \\
		\bottomrule
	\end{tabular}
	
	\vspace*{5mm}
	
	\begin{tabular}{ccccccc}
		\toprule
		{}  &  {}  &  Coefficient  &  Uncertainty  &  \# of Unique  &  \# of Unique  &  {}\\
		$B_{\text{object name}}$  &  Object Type  &  (SLI units)  &  (95\% Conf. Int.)  &  Objects  &  Observations  &  $P > |t|$\\
		\midrule
		IMPROVISED\_1   &   Improvised1 &    $+1.650$ &  $\pm~0.047$ &  13 &  1,014 &  $<1e-3$ \\
		MINE\_2              &     Mine1 &    $+1.459$ &  $\pm~0.039$ &  29 &  1,941 &  $<1e-3$ \\
		OBJECT\_1           &       Object1 &    $+1.070$ &  $\pm~0.056$ &  10 &    613 &  $<1e-3$ \\
		OBJECT\_2             &       Object1 &   $+1.007$ &  $\pm~0.052$ &   9 &    760 &  $<1e-3$ \\
		ARTILLERY\_1  &         Artillery &    $+0.713$ &  $\pm~0.047$ &  19 &  1,019 &  $<1e-3$ \\
		IMPROVISED\_2      &    Improvised2 &  $+0.467$ &  $\pm~0.051$ &  10 &    794 &  $<1e-3$ \\
		MINE\_4            &     Mine1 &   $+0.387$ &  $\pm~0.076$ &   6 &    289 &  $<1e-3$ \\
		MINE\_1           &  Mine2 &   $+0.302$ &  $\pm~0.046$ &  16 &  1,148 &  $<1e-3$ \\
		ARTILLERY\_2 &         Artillery &    $+0.229$ &  $\pm~0.040$ &  28 &  1,831 &  $<1e-3$ \\
		ARTILLERY\_3   &         Artillery &    $+0.145$ &  $\pm~0.076$ &   7 &    302 &  $<1e-3$ \\
		MINE\_5              &  Mine2 &   $+0.013$ &  $\pm~0.039$ &  29 &  1,990 &  $0.512$ \\
		IMPROVISED\_3              &    Improvised2 &  $-0.015$ &  $\pm~0.035$ &  59 &  3,293 &  $0.390$ \\
		IMPROVISED\_4             &    Improvised2 &  $-0.144$ &  $\pm~0.041$ &  24 &  1,645 &  $<1e-3$ \\
		DISC\_1              &             Discs &    $-0.450$ &  $\pm~0.111$ &   9 &    126 &  $<1e-3$ \\
		MINE\_3            &    Mine3 &  $-0.673$ &  $\pm~0.042$ &  22 &  1,485 &  $<1e-3$ \\
		OBJECT\_3             &    Object2  &   $-0.869$ &  $\pm~0.043$ &  20 &  1,425 &  $<1e-3$ \\
		OBJECT\_4           &    Object2 &   $-0.921$ &  $\pm~0.082$ &   5 &    249 &  $<1e-3$ \\
		OBJECT\_5               &    Object2 &   $-1.284$ &  $\pm~0.041$ &  32 &  1,744 &  $<1e-3$ \\
		MINE\_6            &    Mine3 &  $-1.288$ &  $\pm~0.091$ &   6 &    192 &  $<1e-3$ \\
		DISC\_2              &             Discs &    $-1.603$ &  $\pm~0.112$ &   9 &    124 &  $<1e-3$ \\
		\bottomrule
	\end{tabular}
	
	\vspace*{5mm}
	
\end{center}
\newpage

\section{Model 2 Regression Parameters}
\label{appendix:Model2}
\begin{center}
	
	\vspace*{5mm}

	\begin{tabular}{cccccc}
		\toprule
		{}  &  Coefficient  &  Uncertainty  &  \# of Unique  &  \# of Unique  &  {}\\
		$C_{\text{array}}$  &  (SLI units)  &  (95\% Conf. Int.)  &  Objects  &  Observations  &  $P > |t|$\\
		\midrule
			Array 3   &  $+0.277$ &  $\pm~0.015$ &  448 &  9,142 &  $<1e-3$ \\
			Array 1  &  $+0.189$ &  $\pm~0.023$ &  427 &  2,532 &  $<1e-3$ \\
			Array 4 &  $+0.069$ &  $\pm~0.016$ &  415 &  6,996 &  $<1e-3$ \\
			Array 5   &  $-0.246$ &  $\pm~0.019$ &  381 &  4,095 &  $<1e-3$ \\
			Array 2  &  $-0.290$ &  $\pm~0.022$ &  430 &  2,878 &  $<1e-3$ \\
		\bottomrule
	\end{tabular}
	
	\vspace*{5mm}
	
	\begin{tabular}{cccccc}
		\toprule
		{}  &  Coefficient  &  Uncertainty  &  \# of Unique  &  \# of Unique  &  {}\\
		$E_{\text{object type}}$  &  (SLI units)  &  (95\% Conf. Int.)  &  Objects  &  Observations  &  $P > |t|$\\
		\midrule
			Improvised1  &  $+1.610$ &  $\pm~0.041$ &  17 &  1,242 &  $<1e-3$ \\
			Mine1    &  $+1.225$ &  $\pm~0.033$ &  50 &  3,124 &  $<1e-3$ \\
			Object1      &  $+0.808$ &  $\pm~0.036$ &  29 &  1,804 &  $<1e-3$ \\
			Artillery        &  $+0.258$ &  $\pm~0.033$ &  65 &  3,400 &  $<1e-3$ \\
			Improvised2   &  $-0.064$ &  $\pm~0.028$ &  96 &  5,883 &  $<1e-3$ \\
			Mine2 &  $-0.125$ &  $\pm~0.030$ &  66 &  4,089 &  $<1e-3$ \\
			Mine3   &  $-0.677$ &  $\pm~0.037$ &  28 &  1,677 &  $<1e-3$ \\
			Discs            &  $-0.782$ &  $\pm~0.084$ &  18 &    250 &  $<1e-3$ \\
			Object2  &  $-0.917$ &  $\pm~0.031$ &  61 &  3,569 &  $<1e-3$ \\
			Object3   &  $-0.947$ &  $\pm~0.058$ &  18 &    549 &  $<1e-3$ \\
		\bottomrule
	\end{tabular}
	
	\vspace*{5mm}
	
	\begin{tabular}{cccccc}
		\toprule
		{}  &  Coefficient  &  Uncertainty  &  \# of Unique  &  \# of Unique  &  {}\\
		$D(z)$ &  (SLI units)  &  (95\% Conf. Int.)  &  Objects  &  Observations  &  $P > |t|$\\
		\midrule
			$D_{1}$  &  $+2.965$ &  $\pm~0.028$ &  422 &  23,977 &  $<1e-3$ \\
			$D_{2}$ &  $+1.397$ &  $\pm~0.050$ &   68 &   1,665 &  $<1e-3$ \\
		\bottomrule
	\end{tabular}
	
	\vspace*{5mm}

	\begin{tabular}{cccccc}
		\toprule
		{}  &  Coefficient  &  Uncertainty  &  \# of Unique  &  \# of Unique  &  {}\\
		Area  &  (SLI units / $m^2$)  &  (95\% Conf. Int.)  &  Objects  &  Observations  &  $P > |t|$\\
		\midrule
		$F_{\text{area}}$ &  $+4.007$ &  $\pm~0.301$ &  449 &  25,643 &  $<1e-3$ \\
		\bottomrule
	\end{tabular}
	
	\vspace*{5mm}
	
\end{center}

\twocolumn
\section*{Acknowledgment}

The authors would like to acknowledge Dave Chambers from LLNL and the following students from Biola University for their contributions to this work: Jasper Jiang, Daniel Shen, Andy Van Antwerp, Daniel Monroe, Christian Rim, Matt Teodoro, Janice Tjoa, \& Grace Yoon.
Additionally, the authors are grateful for  support  from  the Office  of  Naval  Research.  This  work  was  performed  under the  auspices  of  the  U.S.  Department  of  Energy  by  Lawrence Livermore  National  Laboratory  under  Contract  DE-AC52-07NA27344.

\ifCLASSOPTIONcaptionsoff
  \newpage
\fi



\bibliographystyle{IEEEtran/bibtex/IEEEtran}

\bibliography{report} 

%

%

%

\begin{IEEEbiography}[{\includegraphics[width=1in,height=1.25in,clip,keepaspectratio]{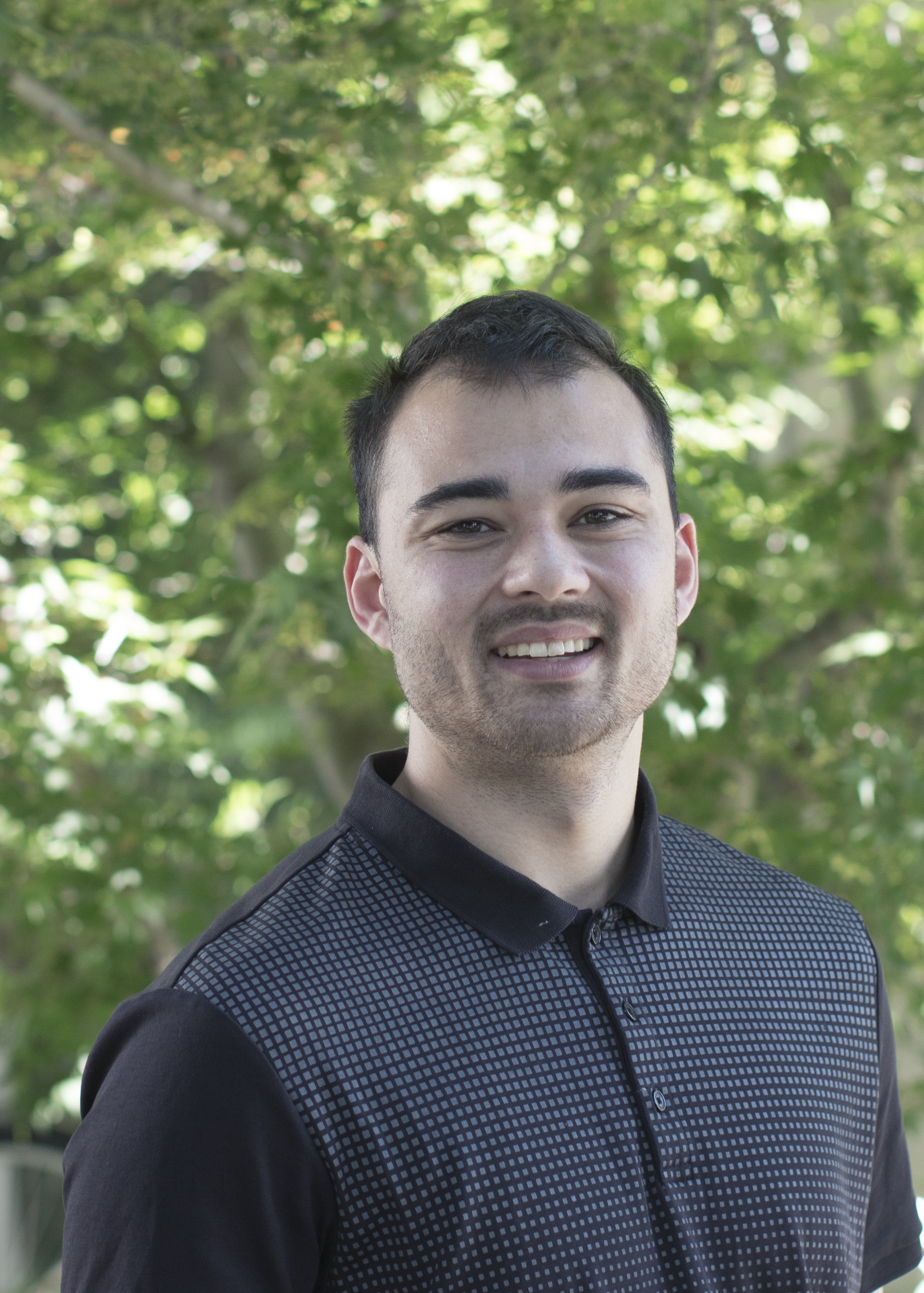}}]{Garrett Stevenson}
is a signal and image processing engineer at LLNL with a background in embedded deep learning, computer vision, nondestructive characterization, and navigation/localization. He received his M.S. in Computer Science at C.S.U. East Bay ('18) with emphases in Computer Vision and Embedded Systems. He currently works in drug discovery and explosives detection using CT images in Ground Penetrating Radar and Checked Baggage. 
\end{IEEEbiography}

\begin{IEEEbiography}[{\includegraphics[width=1in,height=1.25in,clip,keepaspectratio]{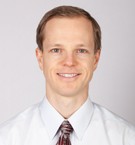}}]{Dr. Jason Wilson}
is the Chair of the Math and Computer Science department and Director of the Quantitative Consulting Center (QCC) at Biola University. Jason is both a Statistician (Ph.D, University of California, Riverside, 2008) and Mathematician (M.A. California State University, Fullerton, 2002). His relevant experiences include a credit card fraud detection project with DHS, statistical training of NASA engineers, and explosives hazard detection in Ground Penetrating Radar with LLNL. 
\end{IEEEbiography}

\begin{IEEEbiography}[{\includegraphics[width=1in,height=1.25in,clip,keepaspectratio]{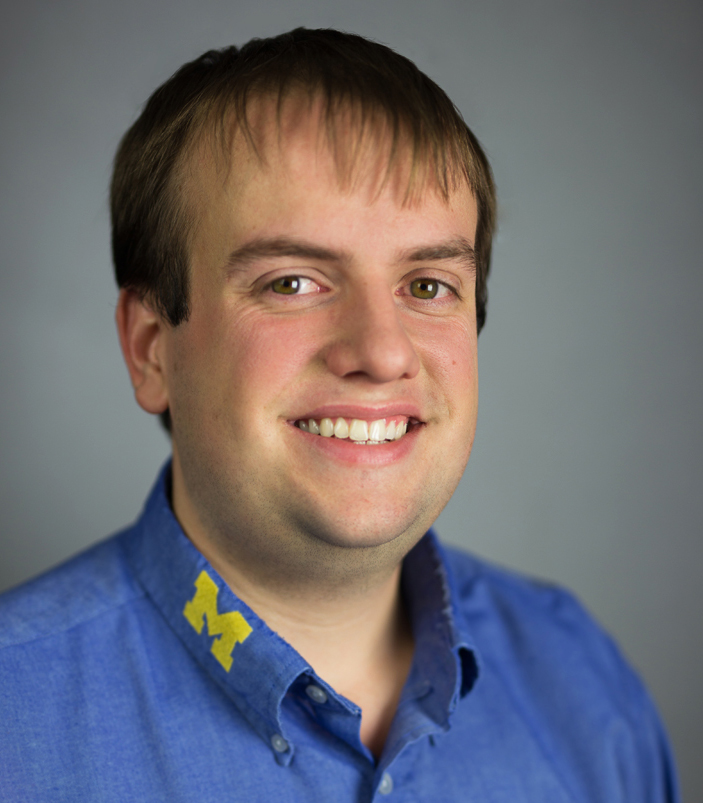}}]{Dr. Brian M. Worthmann}
is a signal and image processing engineer at LLNL with a background in wave-based remote sensing and array signal processing. He received his PhD at the University of Michigan in Applied Physics ('18), where he worked on model-based source localization in ocean acoustics. At LLNL, he has worked with ground penetrating radar signal and image processing, as well as statistical data analysis for hardware and performance characterization.
\end{IEEEbiography}

\begin{IEEEbiography}[{\includegraphics[width=1in,height=1.25in,clip,keepaspectratio]{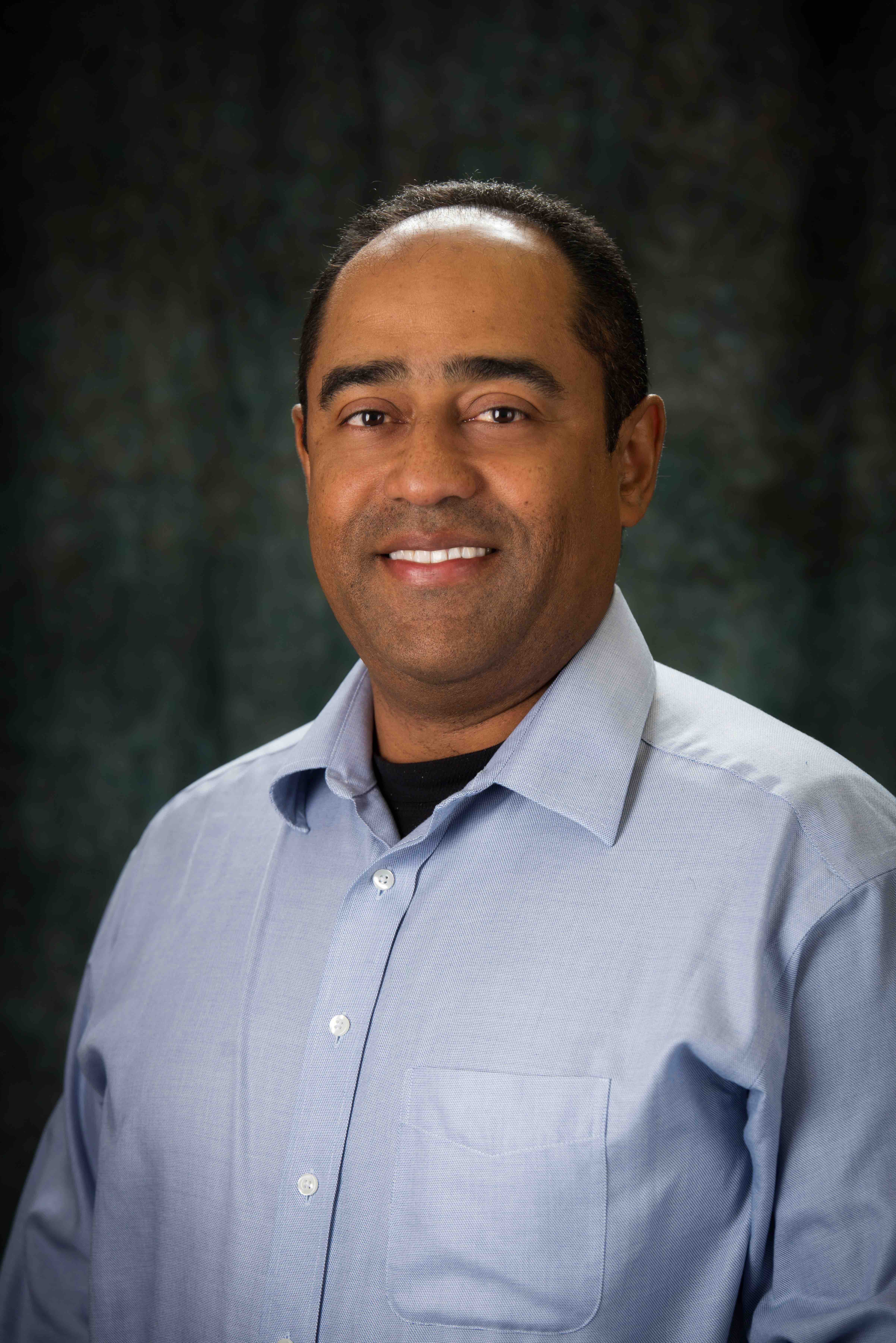}}]{Dr. Wlamir Xavier}
 is an Associate Professor of International Business and Finance and a member of the Quantitative Consulting Center (QCC) at Biola University. Wlamir is a Chemical Engineer (B.S., Instituto Militar de Engenharia, Rio de Janeiro, 1988), Master in Production Engineering (M.S. Santa Catarina Federal University, Florianopolis, Brazil, 1999), and PhD in Business Administration (Univali University, Itajaí, Brazil, 2011). His teaching and research experience includes visiting scholar positions at the University of Paris Dauphine, The University of Pennsylvania, the Copenhagen Business School, the University of La Sabana, Colombia, and Chongqing University, China.  
\end{IEEEbiography}



\end{document}